\documentstyle[12pt]{article}

\begin{document}

\hoffset = -1truecm \voffset = -2truecm \baselineskip = 10 mm

\title{\bf A new modified Altarelli-Parisi evolution equation
with parton recombination in proton}

\author{
{\bf Wei Zhu} and {\bf Jianhong Ruan}\\
\normalsize Department of Physics, East China Normal University,
Shanghai 200062, P.R. China
}
\date{}

\newpage

\maketitle

\vskip 3truecm

\begin{abstract}

	The coefficients of the nonlinear terms in a modified
Altarelli-Parisi evolution equation with parton recombination are
determined in the leading logarithmic ($Q^2$) approximation. The results
are valid in the whole $x$ region and contain the translation
$GG\rightarrow q\overline q$, which is inhibited in the double leading
logarithmic approximation. The comparisons of the new evolution equation
with the Gribov-Levin-Ryskin equation are presented.

\end{abstract}

\newpage

\begin{center}
\section{Introduction}
\end{center}
	
	The inclusion of parton recombination to the proton structure
function through a modified Altarelli-Parisi equation is an interesting
subject involving the test of perturbative QCD and the study of new
effects in small $x$ physics. A traditional tool in this research is the
so-called Gribov-Levin-Ryskin (GLR) equation, presented in
two pioneering papers: one is the idea of shadowing arising from gluon
recombination, which was proposed by Gribov, Levin, Ryskin [1] based on
the AGK (Abramovsky, Gribov, Kancheli) cutting rules [2] in the
double leading logarithmic approximation (DLLA); the other is a
perturbative calculation of the recombination probabilities in the
DLLA by Mueller and Qiu [3], which enables the GLR equation to be
applied phenomenologically.
	
	Unfortunately, a series of questions in the GLR equation
destroyed the gluon recombination effects.  In our previous work [4],
one of us (WZ) has proposed that the application of the AGK cutting rules
is unreasonable in the GLR equation since it sums up the diagrams in which
the cut lines break the important correlation among the initial gluons.
For this reason, a new evolution equation including parton recombination was
established in the leading logarithmic ($Q^2$) approximation (LL($Q^2$)A)
using time ordered perturbation theory (TOPT) instead of the AGK cutting
rules [4]. This equation for parton distributions in the proton can be
generally written as

$$\frac{dG(x_B,Q^2)}{d\ln Q^2}$$
$$=\frac{1}{Q^2}
\int_{(x_1+x_2)\geq x_B} G^{(2)}(x_1, x_2,x_1+\Delta, x_2-\Delta,Q^2)
\sum_iP_i^{GG\rightarrow GG}(x_1,x_2,x_3,x_4,\Delta)$$
$$\delta(x_1+x_2-x_3-x_4)
[\delta(x_3-x_B)+\delta(x_4-x_B)]dx_1dx_2dx_3dx_4d\Delta$$
$$-\frac{2}{Q^2}
\int_{x_1\geq x_B} G^{(2)}(x_1, x_2, x_1+\Delta, x_2-\Delta ,Q^2)
\sum_iP_i^{GG\rightarrow GG}(x_1,x_2,x_3,x_4,\Delta)$$
$$\delta(x_1+x_2-x_3-x_4)
[\delta(x_3-x_B)+\delta(x_4-x_B)]dx_1dx_2dx_3dx_4d\Delta,
\eqno(1a)$$
and

$$\frac{dS(x_B,Q^2)}{d\ln Q^2}$$
$$=\frac{1}{Q^2}
\int_{(x_1+x_2)\geq x_B} G^{(2)}(x_1, x_2,x_1+\Delta, x_2-\Delta,Q^2)
\sum_iP_i^{GG\rightarrow q\overline{q}}(x_1,x_2,x_3,x_4,\Delta)$$
$$\delta(x_1+x_2-x_3-x_4)
[\delta(x_3-x_B)+\delta(x_4-x_B)]dx_1dx_2dx_3dx_4d\Delta$$
$$-\frac{2}{Q^2}
\int_{x_1\geq x_B} G^{(2)}(x_1, x_2, x_1+\Delta, x_2-\Delta ,Q^2)
\sum_iP_i^{GG\rightarrow q\overline{q}}(x_1,x_2,x_3,x_4,\Delta)$$
$$\delta(x_1+x_2-x_3-x_4)
[\delta(x_3-x_B)+\delta(x_4-x_B)]dx_1dx_2dx_3dx_4d\Delta,
\eqno(1b)$$
where we temporarily neglect the linear terms in the Altarelli-Parisi
equation and $G^{(2)}$ is the correlation function for four-gluon lines.
The new equation provides the following physical picture for
the gluon recombination in a QCD evolution process:
the two-parton-to-two-parton $(2 \to 2)$ amplitudes (Figs. 1a-1b) lead to
positive (antiscreening) effects and the interference
amplitudes between the one-parton-to-two-parton $(1 \to 2)$ and the
three-parton-to two-parton $(3 \to 2)$ (Figs. 1c-1f) amplitudes lead to
negative (screening) effects, respectively.  The TOPT-analysis shows that
the above two kinds of amplitudes correspond to the same recombination
function $\sum_iP_i$ but in different kinematic ranges. On the other hand,
the complete corrections of the gluon recombination to the evolution equation
should include the contributions from the virtual diagrams corresponding to
$\delta(x_1-x_B)$ and $\delta(x_2-x_B)$, however, they cancel each
other in the proton [4].

	The quantitative predictions of Eq. (1) depend on the recombination
functions. Mueller and Qiu [3] have derived similar transition
probabilities $\sum_iP_{MQ,i}$ in the DLLA, where only the gluon ladder
graphs are kept at small $x$. We should remember that the AGK cutting
rules are applicable only in diagrams which consist of
gluon ladders [1]. Therefore, the DLLA is a necessary approximation
for using the AGK cutting rules in the GLR equation. On the other hand,
in the Mueller-Qiu approach [3], the cut vertex technique [5] was used to
calculate the transition probabilities, which are entangled in a complex
cut diagram with four-parton propagators [3]. Therefore, the DLLA
is also a convenient approximation in Ref.~[3].

	However, we shall point out that the DLLA leads to a difficulty in
the GLR equation: the transition of gluon$\rightarrow$ quarks is suppressed
in the DLLA-manner. Although the special techniques are used to include the
corrections of gluon recombination to the quark distributions
in Ref.~[3], however, it seems inconsistent in the theory as we will show
in Sect. 5.

	The purpose of this paper is to derive the recombination functions
of Eq. (1) in the LL($Q^2$) approximation. We can generalize the
recombination function to the whole $x$ region and even include the
processes with quarks, since Eq. (1) is not restricted by the AGK
cutting rules [4]. Furthermore, we used TOPT instead of the cut vertex method
in the derivation of Eq. (1) [4]. Thus, we can separately calculate
the simple two-parton-to-two-parton $(2 \to 2)$ processes in the recombination function.
Following the above mentioned derivation, we shall complete
a new modified Altarelli-Parisi evolution equation with parton recombination
in the LL($Q^2$) approximation, which is valid in the whole $x$ region.
Momentum conservation is naturally restored due to the coexistence
of the shadowing- and antishadowing-effects in the new evolution equation.
We also find that the DLLA is a bad approximation even for the
recombination function of gluons in the small-$x$ region.

	The outline of the paper is as follows.
In Sect. 2 we shall derive a set of complete recombination functions.
An undetermined factor in Eq. (1) is the correlation function
$G^{(2)}$, which relates to the nonperturbative structure of the proton.
We use a simple assumption to model $G^{(2)}$ in Sect. 3.
Combining the results of the above two sections, we establish a modified
Altarelli-Parisi equation including parton recombination in Sect. 4.
In Sect. 5 we compare our new equation with the GLR equation.

\newpage
\begin{center}
\section{Recombination functions}
\end{center}
	
	The recombination function in Eq. (1) is factorized due to the
application of TOPT in [4]. All processes of two-parton-to-two-parton
$(2 \to  2 )$ type can be precisely calculated using the
standard perturbative QCD, except that the energies of partons are not
conserved at the vertex connecting with the probe.

	At first, we calculate the recombination function for the
two-gluon-to-two-gluon process. The momenta of all initial and
final partons are on-shell and they are parametrized as (see Fig. 2),

$$p_1=\left [x_1p,\b{0},x_1p\right ]; \hspace{0.3cm}
p_2=\left [x_2p,\b{0},x_2p\right ],$$
$$p_1'=\left [x_1'p,\b{0},x_1'p \right ]; \hspace{0.3cm}
p_2'=\left [x_2'p,\b{0},x_2'p \right ],$$
$$k=\left [x_3p+\frac{l^2_{\perp}}{2x_3p},l_{\perp},x_3p \right ];
\hspace {0.3cm}
l'=\left [x_4p+\frac{l^2_{\perp}}{2x_4p},-l_{\perp},x_4p \right ].
\eqno(2)$$

	We take the physical axial gauge and the light-like vector
$n$ fixes the gauge as $n\cdot A=0$, $A$ being the gluon field.
The corresponding recombination function is defined by [4]

$$\alpha_s^2P_i^{GG\rightarrow GG}(x_1,x_2,x'_1,x'_2,x_3,x_4)dx_4
\frac{dl^2_{\perp}}{l^4_{\perp}}$$
$$=\frac{E_k}{\sqrt{E_{p_1}+E_{p_2}}\sqrt{E_{p'_1}+E_{p'_2}}}
[M(p_1p_2\rightarrow kl')[M(p'_1p'_2\rightarrow kl')]^*]_i
(\frac{1}{E_{p_1}+E_{p_2}-E_k-E_{l'}})^2$$
$$(\frac{1}{2E_k})^2\frac{d^3l'}{(2\pi)^32E_{l'}}$$
$$=\frac{1}{16\pi^2}\frac{x_3x_4}{(x_1+x_2)^3}
[M(p_1p_2\rightarrow kl')[M_(p'_1p'_2\rightarrow kl')]^*]_i
dx_4\frac{dl^2_{\perp}}{l^4_{\perp}}.
\eqno(3)$$

	The index $i$ in Eq. (3) implies the $t$-, $u$-, $s$-channels and
their interference terms. We begin with the $t$-channel graph (Fig. 3).
The contribution of this process to the invariant amplitude is

$$\left\vert M_tM^*_t\right\vert_{x_3=x_B}$$
$$=\frac{g^4}{4}\frac{C_A^2}{N^2-1}\frac{1}{l_L^2}\frac{1}{l_R^2}
\Gamma^{\kappa\rho}C^{\kappa\mu\xi}C^{\rho\phi\alpha}\Gamma^{\lambda\sigma}
C^{\lambda\eta\nu}C^{\sigma\beta\chi}\delta_{\perp}^{pq}\delta_{\perp}^{rs}
[\delta^{ij}-\frac{k^ik^j}{\left\vert\vec{k}\right\vert^2}]
[\delta^{lm}-\frac{l'^ll'^m}{\left\vert\vec{l'}\right\vert^2}],
\eqno(4)$$
where $\Gamma^{\kappa\rho}=g^{\kappa\rho}-\frac{l_L^\kappa n_{\rho}+l_L^\rho n_{\kappa}}{l_L\cdot
n}$;
$(i,j), (l,m), (p,q)$ and $(r,s)$ are the space-indices corresponding
to ($\mu$,$\nu$), ($\alpha$,$\beta$), ($\xi$,$\eta$) and ($\phi$,$\chi$),
respectively. We need to distinguish the probing place, for example,
the transfer momenta $l_L$ and $l_R$ are determined by two down-vertices
in Fig. 3 where the probing place is $x_3=x_B$,
 i.e.,

$$l_L=\left [(x_4-x_2)p+\frac{l^2_{\perp}}{2x_4p},-l_{\perp},(x_4-x_2)p \right ],$$
$$l_R=\left [(x_4-x'_2)p+\frac{l^2_{\perp}}{2x_4p},-l_{\perp},(x_4-x'_2)p \right ],
\eqno(5)$$ since energy is not conservation at other vertices.

The algebra in Eq. (4) can be straightforwardly derived without any
approximation in computer. The contributions from the $u$-channel and
the interference channels can be similarly obtained by using the interchanges
of the corresponding momenta.

	Now we turn to discuss the $s$-channel as shown in Fig. 4.
We should note that the massless partons with the parallel
momenta can go on-mass-shell simultaneously in the collinear case
and the collinear singularity may arise in the $s$-channel since
$l^2=(p_1+p_2)^2=(p_1'+p_2')^2=0$. Fortunately, we now have an useful tool
[6,7] to pick up the short-distance contributions in the propagator with
collinear divergence: we use the following special propagators (for quarks)

$$S(l)=\frac{\gamma\cdot n}{2l\cdot n}, \eqno(6)$$
and (for gluons)

$$G^{\kappa\rho}(l)
=\frac{n^\kappa n^\rho}{(l\cdot n)^2}, \eqno(7)$$
to replace the normal Feynman propagators, respectively. Using the
definitions

$$\overline{n}^{\mu}\equiv (\overline{n}^0, \overline{n}_\perp,
\overline{n}^3)=\frac{1}{\sqrt{2}}(1, 0_\perp,1);\hspace{0.3cm}
n^{\mu}=\frac{1}{\sqrt{2}}(1, 0_\perp, -1),\eqno(8)$$
or equivalently

$$\overline{n}^{\mu}\equiv(\overline{n}^+,
\overline{n}^-,\overline{n}_\perp)
=(1, 0, 0_\perp);\hspace{0.3cm} n^{\mu}=(0,1, 0_\perp), \eqno(9)$$
we have $\overline{n}^2=0, n^2=0$, and $\overline{n}\cdot n=1$.
The contribution of the $s$-channel (Fig. 4) to the invariant amplitude
can then be safely calculated as,

$$\left\vert M_sM^*_s\right\vert$$
$$=\frac{g^4}{4}\frac{C_A^2}{N^2-1}(\frac{1}{l\cdot n})^4
n^{\kappa}C^{\kappa\xi\phi}n^{\lambda}C^{\lambda\eta\chi}
n^{\rho}C^{\rho\alpha\mu}n^{\sigma}C^{\sigma\beta\nu}\delta_{\perp}^{pq}\delta_{\perp}^{rs}
[\delta^{ij}-\frac{k^ik^j}{\left\vert\vec{k}\right\vert^2}]
[\delta^{lm}-\frac{l'^ll'^m}{\left\vert\vec{l'}\right\vert^2}]$$
$$=g^4\frac{C_A^2}{N^2-1}\frac{(x_4-x_3)^2(x_2-x_1)(x_2'-x_1')}{2x^4},
\eqno(10)$$

where we have used the following replacement in the calculations:

$$n^{\kappa}C^{\kappa\xi\phi}=n^{\kappa}(p_1-p_2)^{\kappa}. \eqno(11)$$

Obviously, the contributions given by the $s$-channel and corresponding
interference channels integrate to zero if the correlation function
$G^{(2)}(x_1,x_2,x_1',x_2,Q^2)$ is symmetric
under $x_1\leftrightarrow x_2$ and $x_1'\leftrightarrow x_2'$, since these
contributions contain the factor $(x_2-x_1)$ or $(x_2'-x_1')$. Thus,
we can forget the $s$-channel and its interference terms in $GG\rightarrow GG$.
The contributions of Feynman diagram with four-gluon vertex are neglected
because they are less singular. The total contributions of $GG\rightarrow GG$ are given
in Appendix.

	The recombination function $\sum_iP_i^{GG\rightarrow GG}$ has poles
at $x_1=0$ and $x_2=0$ (see Eq. (A.1)).  However, Eq. (1.a) is really
infrared safe. One can simply check this conclusion as follows. For example,
$x_1=0$ implies that $\Delta=0$ since only twist-4 or twist-2 amplitudes
contribute to an unpolarized amplitude. Summing over the contributions of
the right-hand side of Eq. (1.a) at the poles, one can find that

$$\int_{1\geq x_1\geq x_B} G^{(2)}(x_1=0, x_2, \Delta=0, Q^2)
\sum_iP_i^{GG\rightarrow GG}(x_1=0,x_2,\Delta=0,x_3,x_4,\Delta)$$
$$\delta(x_1+x_2-x_3-x_4)
[\delta(x_3-x_B)+\delta(x_4-x_B)]dx_2dx_3dx_4$$
$$+\int_{1\geq x_1\geq x_B} G^{(2)}(x_1, x_2=0, \Delta=0, Q^2)
\sum_iP_i^{GG\rightarrow GG}(x_1,x_2=0,\Delta=0,x_3,x_4,\Delta)$$
$$\delta(x_1+x_2-x_3-x_4)
[\delta(x_3-x_B)+\delta(x_4-x_B)]dx_1dx_3dx_4$$
$$-2\int_{1\geq x_1\geq x_B} G^{(2)}(x_1, x_2=0, \Delta=0,Q^2)
\sum_iP_i^{GG\rightarrow GG}(x_1,x_2=0,\Delta=0,x_3,x_4,\Delta)$$
$$\delta(x_1+x_2-x_3-x_4)
[\delta(x_3-x_B)+\delta(x_4-x_B)]dx_1dx_3dx_4=0,
\eqno(12)$$
where the two positive terms are symmetric under $x_1\leftrightarrow x_2$
and $x_1\ne 0$ in the negative interference term. Therefore, Eq. (1)
is infrared safe.

Furthermore, unlike the GLR equation, the momentum is conserved
in Eq. (1), i.e.,

$$\frac{d\int^1_0dx_Bx_BG(x_B,Q^2)}{d\ln Q^2}=0, \eqno(13)$$
and

$$\frac{d\int^1_0dx_Bx_Bq(x_B,Q^2)}{d\ln Q^2}=0. \eqno(14)$$

	Now let us return to consider the recombination functions for
$GG\rightarrow q\overline{q}$. For example, the contribution of the
$t$-channel process in Fig. 5 to the invariant amplitude is

$$\left\vert M_tM^*_t\right\vert_{x_3=x_B}$$
$$=\frac{g^4}{4}\frac{T_f}{N(N^2-1)} \frac{x_4^2}{x_2x_2'}\frac{1}{l^4_\perp}
Tr[\gamma\cdot k\gamma^{\eta}\gamma\cdot l_R\gamma^{\chi}\gamma\cdot l'
\gamma^{\phi}\gamma\cdot l_L\gamma^{\xi}]
\delta_{\perp}^{pq}\delta_{\perp}^{rs}.
\eqno(15)$$

The final results for $\sum_i^{GG\rightarrow q\overline{q}}$ are listed in
Appendix, where the contributions of the $s$-channel and corresponding
interference channels vanish due to the same symmetry as in
$\sum_iP_i^{GG\rightarrow GG}$.

We need only consider the gluon recombination at small $x$,
since gluons dominate the small-$x$ behavior of the parton distributions
in proton. However, in principle, our equation can include the
recombination of quark-quark, quark-antiquark and quark-gluon in the LLA.
For this end, we list all recombination functions in Appendix.
We will discuss the properties of our results in Sect. 5.

\newpage
\begin{center}
\section{Parton correlation functions}
\end{center}

	Equation (1) includes the gluon correlation function
$G^{(2)}(x_1,x_2;x'_1,x'_2,Q^2)$. In general, the parton density is a
concept defined at twist-2. The parton correlation function
is a generalization of the parton density beyond the leading twist and
it has not yet been determined both in theory and experiment.
For comparison with the GLR equation, we shall use a toy model as in
Ref.~[3].  First of all, we assume that

$$G^{(2)}(x_1, x_2,x_1+\Delta, x_2-\Delta,Q^2)$$
$$\rightarrow G^{(2)}(x_1, x_2,x_1+\Delta, x_2-\Delta,Q^2)\delta(x_1-x_2)
\delta(\Delta). \eqno(16)$$
One can understand Eq. (16) as follows. Parton recombination is expected to
be significant only when the actual spatial overlap between the
fusing partons becomes sufficiently large in the interaction time $\tau_{int}$
of probe-target both in the transverse- and longitudinal-directions. This means that
(i) the transverse area occupied by partons should become compared to the
effective transverse area; (ii) only two partons with the same value of
rapidity $y=\ln{\frac{1}{x}}$ and same impact parameter
can overlap their wave functions in the longitudinal direction during
$\tau_{int}$, since the initial partons are parallel moving along the
$z$-direction. Therefore, the fusion between the gluons with different values
of $y$ is suppressed.

	We consider that the different branches of the parton cascade evolve
independently; this is valid in the large $N_c$ limit [8]. Thus,
$G^{(2)}(x,Q^2)\sim G^2(x,Q^2)$, where $G(x,Q^2)$ is the usual gluon density.
However, the density $G(x,Q^2)$ can not be normalized since
$G(x,Q^2)\sim \infty$ when $x\rightarrow 0$.  A more reasonable approach is
to use the gluon density in rapidity space $xG(x,Q^2)=dn_G/dy$ instead of
$G(x,Q^2)$. The normalization condition is

$$\sum_a\frac {dn_a}{dy}=1, \eqno(17)$$
due to momentum conservation, where the sum is over all parton flavors.
Therefore, we can assume the two-gluon distribution per unit area in
proton to be

$$(x_1+x_2)G^{(2)}(x_1,x_2,\Delta, Q^2)=\frac{1}{4\pi R^2}[x_1G(x_1,Q^2)][x_2G(x_2,Q^2)]
\delta(x_1-x_2)\delta(\Delta),
\eqno(18)$$
where the factor is from a normalization factor of the parton
correlation function [3,6].  Although Eq. (18) is a phenomenological
model, however, its $x$-dependence can be checked by experiment.

\newpage
\begin{center}
\section{Modified Altarelli-Parisi equation}
\end{center}

	In this section we derive a modified Altarelli-Parisi equation
with gluon recombination. Inputting Eq. (18) with the recombination functions
to Eq. (1) and adding the linear terms corresponding to the Altarelli-Parisi
equation, we have

$$\frac{dx_BG(x_B,Q^2)}{d\ln Q^2}$$
$$=\frac{C_A\alpha_s}{\pi}\int^1_{x_B}\frac{dx_1}{x_1}\frac{x_B}{x_1}
P^{G\rightarrow G}_{AP}(x_1,x_B)x_1G(x_1,Q^2)$$
$$+\frac{\alpha_s^2(Q^2)}{4\pi R^2Q^2}
\int_{x_B/2}^{x_B}dx_1x_Bx_1G^2(x_1, Q^2)
\sum_iP_i^{GG\rightarrow G}(x_1,x_B)$$
$$-\frac{\alpha_s^2(Q^2)}{4\pi R^2Q^2}
\int_{x_B}^{1/2}dx_1x_Bx_1G^2(x_1, Q^2)
\sum_iP_i^{GG\rightarrow G}(x_1,x_B),
\eqno(19.a)$$
where the recombination functions become simple after
integral over $x_2$, $\Delta$, $x_3$ and $x_4$ under the assumption (18):

$$\sum_iP_i^{GG\rightarrow G}(x_1,x_B)$$
$$
=\frac{9}{64}\frac{(2x_1-x_B)(72x_1^4-48x_1^3x_B+140x_1^2x_B^2-116x_1x^3_B+29x_B^4)}{x_1^5x_B}, \eqno(19.b)$$
which is valid in the whole $x$ region. The second term on the right-hand side
of Eq. (19.a) is positive due to the contributions of Figs. 1a-1b and
is called as antishadowing effect. The negative term in Eq. (19.a) arises
from the shadowing corrections from the interference diagrams Figs. 1.c-1.f.

	We have a similar result for $GG\rightarrow q\overline {q}$:

$$\frac{dx_BS(x_B,Q^2)}{d\ln Q^2}$$
$$=\frac{2T_f\alpha_s}{\pi}\int^1_{x_B}\frac{dx_1}{x_1}\frac{x_B}{x_1}
P^{G\rightarrow q}_{AP}(x_1,x_B)x_1G(x_1,Q^2)$$
$$+\frac{\alpha_s^2(Q^2)}{4\pi R^2Q^2}
\int_{x_B/2}^{x_B}dx_1x_Bx_1G^2(x_1, Q^2)
\sum_iP_i^{GG\rightarrow q}(x_1,x_B)$$
$$-\frac{\alpha_s^2(Q^2)}{4\pi R^2Q^2}
\int_{x_B}^{1/2}dx_1x_Bx_1G^2(x_1, Q^2)
\sum_iP_i^{GG\rightarrow q}(x_1,x_B),
\eqno(19.c)$$
where the recombination function with the assumption (18) is

$$\sum_iP_i^{GG\rightarrow q}(x_1,x_B)$$
$$=\frac {1}{96}\frac {
(2x_1-x_B)^2
(18x_1^2-21x_1x_B+14x_B^2)}
{x_1^5}. \eqno(19.d)$$

We will discuss the properties of our new modified Altarelli-Parisi
equation (19) in the next section.

\newpage
\begin{center}
\section{Discussions}
\end{center}

	In order to compare Eq. (19) with the GLR equation,
we begin with a review of the DLLA. As definited by the Altarelli-Parisi
equation, the DLLA means that in each order in $\alpha_s$, one keeps only
the $\ln(Q^2/\mu^2)\ln(1/x)$ factor in the solutions of the evolution
equation, or equivalently, only the terms having $1/z=x_1/x_B$ factor
in the splitting function are necessary to generate larger logarithms
in $x_B$ (Fig. 6). We know that the splitting functions
$P_{AP}^{G\rightarrow G}(z)$ and $P_{AP}^{q\rightarrow G}(z)$ have the
lowest $z$-power ($\sim 1/z$), while $P_{AP}^{q\rightarrow q}(z)$ and
$P_{AP}^{G\rightarrow q}(z)$ vanish in the DLLA. Therefore,
the DLLA diagram consists of the gluon ladders, since any transition of
gluon$\rightarrow$ quark breaks the ladder-structure. Similarly, we also
find (see the Appendix) that only
$P_i^{GG\rightarrow G}$, $P_i^{q\overline{q}\rightarrow G}$
and $P_i^{qG\rightarrow G}$ give the leading contributions in the DLLA
since they have the factor $1/z=(x_1+x_2)/x_B$.
Thus, we can conclude that any transition of $G\rightarrow q$ or
$GG\rightarrow q$ is suppressed in the DLLA-approach.

	Now let us remember the GLR equation, which has following form
in [3]:

$$\frac{dx_BG(x_B,Q^2)}{d\ln Q^2}$$
$$=\frac{3\alpha_s}{\pi}\int^1_{x_B}dx_1G(x_1,Q^2)$$
$$-5.05(\frac{\alpha_s}{RQ})^2\int_{x_B}^{x_0}
\frac{dx_1}{x_1}[x_1G(x_1, Q^2)]^2, \eqno(20.a)$$
and

$$\frac{dx_BS(x_B,Q^2)}{d\ln Q^2}$$
$$=\frac{\alpha_s}{4\pi}\int^1_{x_B}\frac{dx_1}{x_1}\frac{x_B}{x_1}
x_1G(x_1,Q^2)$$
$$-0.17\frac{\alpha_s^2(Q^2)}{160R^2Q^2}[x_BG(x_B,Q^2)]^2-
0.32\frac{\alpha_s(Q^2)}{Q^2}\int_{x_B}^{x_0}\frac{dx_1}{x_1}\frac{x_B}{x_1}
P_{MQ}^{GG\rightarrow q\overline{q}}x_1H(x_1,Q^2), \eqno(20.b)$$
with

$$\frac{dx_1H(x_1,Q^2)}{d\ln Q^2}$$
$$=-5.05(\frac{\alpha_s}{RQ})^2\int_{x_B}^{x_0}
\frac{dz}{z}[zG(z,Q^2)]^2, \eqno(20.c)$$
and

$$P_{MQ}^{GG\rightarrow q\overline{q}}(z)=-2z+15z^2-30z^3+18z^4. \eqno(20.d)$$

As we have emphasized, the DLLA inhibits any transition of
$G\rightarrow q$ or $GG\rightarrow q$. Therefore, Ref.~[3] takes some
special treatments to realize the above mentioned transition. For
example, the last term on the right-hand side
of Eq. (20.b) with Eq. (20.c) comes from a contribution of the next leading
logarithms in $Q^2$ and is inconsistent with the DLLA. On the other hand,
$\delta(x_1-x_B)$ is inserted by hand in the derivation of the second term
of Eq. (20.b). However, $\delta(x_1-x_B)$ implies a cut line through
an initial gluon line with momentum $x_1$ and is really the contribution of the
virtual diagram to $(d/d\ln Q^2)x_BG(x_B,Q^2)$; this contribution
should therefore be canceled in the proton [4].
Even if we forget the above mentioned two questions in Eq. (20),
we still cannot prevent the linear term in Eq. (20) from mixing the
quark line with the gluon ladder at each $Q^2_i$. In consequence, either the
evolution from $Q^2_i$ to $Q^2_i+\Delta Q^2$ will stop in the DLLA-manner
due to the inhibition of gluon$\rightarrow$quark,
or the DLLA as well as the AGK cutting rules, which regard the gluon ladder
as the pomeron [1], can not be applied in the GLR equation.

	Unlike Eq. (20), our new equation (19) includes
the transition of gluon$\rightarrow$quark in the whole $x$ region.
Another important difference between Eqs. (19) and (20) is that the positive
antishadowing effects are separated from the negative shadowing effects in
Eq. (19) because they have different kinematical domains in $x$.

	We emphasize that the DLLA is an unreasonable approximation even
in Eq. (19.a). In fact, we take the DLLA and keep the leading term
$\sim 1/z$ in Eq. (19.a). In this case, this equation can be simplified as

$$\frac{dx_BG(x_B,Q^2)}{d\ln Q^2}$$
$$=\frac{3\alpha_s}{\pi}\int^1_{x_B}dx_1G(x_1,Q^2)$$
$$+1.6(\frac{\alpha_s}{RQ})^2\int_{x_B/2}^{x_B}\frac{dx_1}{x_1}
[x_1G(x_1, Q^2)]^2-1.6(\frac{\alpha_s}{RQ})^2\int_{x_B}^{1/2}
\frac{dx_1}{x_1}[x_1G(x_1, Q^2)]^2, \eqno(21)$$
where the upper-limit of the integral in the negative interfering terms
is $1/2$ due to the restriction $x_1+x_2\leq 1$.
There is such gluon density, for example, $xG(x,Q^2)\sim x^{\lambda_c}
(1-x)^7$, where the net recombination effects disappear
due to the balance of the shadowing- and antishadowing-effects.
The calculation shows that $\lambda_c=1$ and $0.5$ for Eq. (19.a) and
Eq. (21), respectively. Therefore, we can find that the DLLA obviously
distorts the recombination effects even in the gluon evolution equation
at small $x$.

	Finally, we estimate the size of the gluon recombination effect.
On can use

$$W=\frac{nonlinear~terms}{linear~terms}\geq \alpha_s, \eqno(22)$$
to determine the kinematic range, where the nonlinear effects can not be
neglected in Eq. (19). The approximation solutions of Eq. (22) are

$$xG(x,Q^2)\geq\frac{(RQ)^2}{1.6}. \eqno(23)$$

The HERA data show that $xG(x,Q^2)\sim 2-10$ for $Q^2=(1-10)~GeV^2$ and
$x=10^{-3}$. Thus, we can expect that the gluon recombination effects will
be appeared in the proton structure function in the range $Q^2<10~GeV^2$ and
$x<10^{-3}$ if $R<<1$ fm (i.e., the gluon distribution in the proton
has the "hot spots"-structure).
	
	In summary, the coefficients of the nonlinear terms in a modified
Altarelli-Parisi evolution equation with gluon recombination are determined
in the leading logarithmic ($Q^2$) approximation. The results are valid in
the whole x region. The comparisons of the new evolution equation with the
GLR equation are presented.  We expect that the gluon recombination effects
of the new equation can be appeared in the proton structure function
in the HERA-domain of small $x$ and low $Q^2$ if the "hot spots" structure
exists in the proton.

\vspace{0.3cm}

\noindent {\bf Acknowledgments}:

We would like to thank Jianwei Qiu for very helpful discussions and
pointing out Eqs. (6)-(7).
We would also like to acknowledge D. Indumathi for useful
comments. This work was supported by National Natural Science Foundation of
China and `95-Climbing' Plan of China.

\newpage
\begin{center}
\Large {\bf Appendix}
\end{center}
\normalsize

    We summary the complete QCD calculations of the recombination functions as follows.

1.  $GG\rightarrow GG$ (Fig. 7).

$$ M_tM^*_t \vert_{x_3=x_B}
=\frac{g^4}{4}<\frac{9}{8}>
C^{\mu\xi\kappa}\frac{\Gamma^{\kappa\rho}}{l_L^2}C^{\rho\phi\alpha}
C^{\nu\eta\lambda}\frac{\Gamma^{\lambda\sigma}}{l_R^2}C^{\sigma\chi\beta}\delta_{\perp}^{pq}\delta_{\perp}^{rs}
[\delta^{ij}-\frac{k^ik^j}{\left\vert\vec{k}\right\vert^2}]
[\delta^{lm}-\frac{{l'}^l{l'}^m}{\left\vert\vec{l}'\right\vert^2}],
$$

$$ M_uM^*_u\vert_{x_3=x_B}=\frac{g^4}{4}<\frac{9}{8}>C^{\alpha\xi\kappa}
\frac{\Gamma^{\kappa\rho}}{l_L^2}C^{\rho\phi\mu}C^{\beta\eta\lambda}
\frac{\Gamma^{\lambda\sigma}}{l_R^2}
C^{\sigma\chi\nu}\delta_{\perp}^{pq}\delta_{\perp}^{rs}
[\delta^{ij}-\frac{k^ik^j}{\left\vert\vec{k}\right\vert^2}]
[\delta^{lm}-\frac{{l'}^l{l'}^m}{\left\vert\vec{l}'\right\vert^2}],
$$

$$ M_sM^*_s\vert_{x_3=x_B}
=\frac{g^4}{4}<\frac{9}{8}>C^{\xi\phi\kappa} \frac{n_\kappa n_\rho}{(l_L\cdot
n)^2}C^{\rho\alpha\mu}C^{\beta\nu\sigma} \frac{
n_\sigma n_\lambda}{(l_R\cdot n)^2}C^{\lambda\eta\chi}
\delta_{\perp}^{pq}\delta_{\perp}^{rs}
[\delta^{ij}-\frac{k^ik^j}{\left\vert\vec{k}\right\vert^2}]
[\delta^{lm}-\frac{{l'}^l{l'}^m}{\left\vert\vec{l}'\right\vert^2}],
$$

$$ M_tM^*_u\vert_{x_3=x_B}
=\frac{g^4}{4}<\frac{9}{16}>C^{\mu\xi\kappa}
\frac{\Gamma^{\kappa\rho}}{l_L^2}C^{\rho\phi\alpha}C^{\beta\eta\lambda}
\frac{\Gamma^{\lambda\sigma}}{l_R^2}
C^{\sigma\chi\nu}\delta_{\perp}^{pq}\delta_{\perp}^{rs}
[\delta^{ij}-\frac{k^ik^j}{\left\vert\vec{k}\right\vert^2}]
[\delta^{lm}-\frac{{l'}^l{l'}^m}{\left\vert\vec{l}'\right\vert^2}],
 $$

$$ M_tM^*_s\vert_{x_3=x_B}
=\frac{g^4}{4}<\frac{9}{16}>C^{\mu\xi\kappa}
\frac{\Gamma^{\kappa\rho}}{l_L^2}C^{\rho\phi\alpha}C^{\beta\nu\sigma}
\frac{n^{\sigma}n^{\lambda}}{(l_R\cdot n)^2}C^{\lambda\eta\chi}
\delta_{\perp}^{pq}\delta_{\perp}^{rs}$$
$$[\delta^{ij}-\frac{k^ik^j}{\left\vert\vec{k}\right\vert^2}]
[\delta^{lm}-\frac{{l'}^l{l'}^m}{\left\vert\vec{l}'\right\vert^2}],
$$

$$ M_uM^*_s\vert_{x_3=x_B}
=\frac{g^4}{4}<\frac{9}{16}>C^{\mu\xi\kappa}
\frac{\Gamma^{\kappa\rho}}{\l_L^2}C^{\rho\phi\alpha}
C^{\beta\nu\sigma}
\frac{n^{\sigma}n^{\lambda}}{(l_R\cdot n)^2}C^{\lambda\eta\chi}
\delta_{\perp}^{pq}\delta_{\perp}^{rs}$$
$$[\delta^{ij}-\frac{k^ik^j}{\left\vert\vec{k}\right\vert^2}]
[\delta^{lm}-\frac{{l'}^l{l'}^m}{\left\vert\vec{l}'\right\vert^2}],
$$

    We have

$$\alpha_s^2\sum_iP_i^{GG\rightarrow GG}(x_1,x_2,x_3,x_4)\vert_{x_3=x_B}
=\frac{1}{16\pi ^2}\frac{x_3x_4}{(x_1+x_2)^3} [M_tM_t^*
\vert_{x_3=x_B}+M_uM_u^* \vert_{x_3=x_B}$$
$$+ M_sM_s^* \vert_{x_3=x_B}
+2M_tM_u^* \vert_{x_3=x_B}+2M_tM_s^* \vert_{x_3=x_B}+ 2M_uM_s^*
\vert_{x_3=x_B}]$$

and

$$\alpha_s^2\sum_iP_i^{GG\rightarrow GG}(x_1,x_2,x_3,x_4)\vert_{x_4=x_B}
=\frac{1}{16\pi ^2}\frac{x_3x_4}{(x_1+x_2)^3} [M_tM_t^*
\vert_{x_4=x_B}+M_uM_u^* \vert_{x_4=x_B}$$
$$+ M_sM_s^* \vert_{x_4=x_B}
 +2M_tM_u^* \vert_{x_4=x_B}+2M_tM_s^* \vert_{x_4=x_B}+ 2M_uM_s^*
\vert_{x_4=x_B}]$$

    Therefore£¬ the recombination is

$$P^{GG\rightarrow G}(x_1,x_B)$$
$$=\int
\sum_iP_i^{GG\rightarrow GG}(x_1,x_2,x_3,x_4)\delta(x_1+x_2-x_3-x_4)
[\delta(x_3-x_B)+\delta(x_4-x_B)]dx_3dx_4$$
$$=\frac{9}{64}\frac{(2x_1-x_B)(72x_1^4-48x_1^3x_B+140x_1^2x_B^2-116x_1x^3_B+29x_B^4)}{x_1^5x_B}.\eqno(A-1)$$

2. $GG\rightarrow q \overline q$ (Fig.8).

$$ M_tM^*_t\vert_{x_3=x_B}
=\frac{g^4}{4}<\frac{1}{12}> Tr[\gamma\cdot k  \gamma^\eta
\frac{\gamma\cdot l_R}{l_R^2} \gamma^\chi \gamma\cdot l' \gamma^\phi
\frac{\gamma\cdot l_L}{l_L^2} \gamma^\xi]
\delta_{\perp}^{pq}\delta_{\perp}^{rs} $$

$$ M_uM^*_u\vert_{x_3=x_B}
=\frac{g^4}{4}<\frac{1}{12}> Tr[\gamma\cdot k  \gamma^\chi
\frac{\gamma\cdot l_R}{l_R^2} \gamma^\eta \gamma\cdot l' \gamma^\xi
\frac{\gamma\cdot l_L}{l_L^2} \gamma^\phi]
\delta_{\perp}^{pq}\delta_{\perp}^{rs}     $$

$$ M_sM^*_s\vert_{x_3=x_B}
=\frac{g^4}{4}<\frac{3}{16}> Tr[
\gamma\cdot l'  \gamma^{\sigma}\gamma\cdot k  \gamma^{\rho}] \frac{n^{\rho}n^{\kappa}}{(l_L\cdot
n)^2}\frac{n^{\sigma}n^{\lambda}}{(l_R\cdot n)^2} C^{\kappa\xi\phi}
C^{\lambda\eta\chi} \delta_{\perp}^{pq}\delta_{\perp}^{rs} $$

$$ M_tM^*_u\vert_{x_3=x_B}=\frac{g^4}{4}<-\frac{1}{96}> Tr[\gamma\cdot k  \gamma^\chi
\frac{\gamma\cdot l_R}{l_R^2} \gamma^\eta \gamma\cdot l' \gamma^\phi
\frac{\gamma\cdot l_L}{l_L^2} \gamma^\xi]
\delta_{\perp}^{pq}\delta_{\perp}^{rs} $$

$$ M_tM^*_s\vert_{x_3=x_B}=\frac{-ig^4}{4}<\frac{3i}{32}> Tr[
\gamma\cdot l' \gamma^\phi \frac{\gamma\cdot l_L}{l_L^2}\gamma^\xi\gamma\cdot k  \gamma^\sigma]
\frac{n^\sigma n^\lambda}{(l_R\cdot n)^2}C^{\lambda\eta\chi}
\delta_{\perp}^{pq}\delta_{\perp}^{rs}   $$

$$ M_uM^*_s\vert_{x_3=x_B}=\frac{-ig^4}{4}<\frac{3i}{32}> Tr[
\gamma\cdot l' \gamma^\xi \frac{\gamma\cdot l_L}{l_L^2}\gamma^\phi\gamma\cdot k  \gamma^\sigma]
\frac{n^\sigma n^\lambda}{(l_R\cdot n)^2}C^{\lambda\eta\chi}
\delta_{\perp}^{pq}\delta_{\perp}^{rs}   $$

    We have

$$\alpha_s^2\sum_iP_i^{GG\rightarrow q\bar q}(x_1,x_2,x_3,x_4)\vert_{x_3=x_B}
=\frac{1}{16\pi ^2}\frac{x_3x_4}{(x_1+x_2)^3} [M_tM_t^*
\vert_{x_3=x_B}+M_uM_u^* \vert_{x_3=x_B}$$
$$+ M_sM_s^* \vert_{x_3=x_B}
 +2M_tM_u^* \vert_{x_3=x_B}+2M_tM_s^* \vert_{x_3=x_B}+ 2M_uM_s^*
\vert_{x_3=x_B}]$$

and

$$\alpha_s^2\sum_iP_i^{GG\rightarrow q\bar q}(x_1,x_2,x_3,x_4)\vert_{x_4=x_B}
=\frac{1}{16\pi ^2}\frac{x_3x_4}{(x_1+x_2)^3} [M_tM_t^*
\vert_{x_4=x_B}+M_uM_u^* \vert_{x_4=x_B}$$
$$+ M_sM_s^* \vert_{x_4=x_B}
 +2M_tM_u^* \vert_{x_4=x_B}+2M_tM_s^* \vert_{x_4=x_B}+ 2M_uM_s^*
\vert_{x_4=x_B}]$$

    Therefore, the recombination is

$$P^{GG\rightarrow
q}(x_1,x_B)=P^{GG\rightarrow \overline{q}}(x_1,x_B)$$
$$=\int
\sum_iP_i^{GG\rightarrow q\bar q}(x_1,x_2,x_3,x_4)
\delta(x_1+x_2-x_3-x_4)
\delta(x_3-x_B)dx_3dx_4$$
$$=\frac{1}{96}\frac{(2x_1-x_B)^2(18x_1^2-21x_1x_B+14x_B^2)}{x_1^5}
.\eqno(A-2)$$

3.  $qq\rightarrow qq$ (Fig.9).

$$ M_tM^*_t\vert_{x_3=x_B}
=\frac{g^4}{4}<\frac{2}{9}> Tr[ \gamma\cdot
p_1 \gamma^\mu\gamma\cdot k  \gamma^\nu] Tr[ \gamma\cdot p_2
\gamma^\alpha\gamma\cdot l'  \gamma^\beta]
\frac{\Gamma^{\mu\alpha}}{l_L^2}\frac{\Gamma^{\nu\beta}}{l_R^2} $$

$$ M_uM^*_u\vert_{x_3=x_B}=\frac{g^4}{4}<\frac{2}{9}>Tr[
\gamma\cdot p_1 \gamma^\mu\gamma\cdot l'  \gamma^\nu] Tr[
\gamma\cdot p_2 \gamma^\alpha\gamma\cdot k  \gamma^\beta]
\frac{\Gamma^{\mu\alpha}}{l_L^2}\frac{\Gamma^{\nu\beta}}{l_R^2}
$$

$$ M_tM^*_u\vert_{x_3=x_B}=\frac{g^4}{4}<-\frac{2}{27}> Tr[\gamma\cdot
p_1 \gamma^\mu\gamma\cdot k  \gamma^\beta
\gamma\cdot p_2 \gamma^\alpha \gamma\cdot l'  \gamma^\nu ]
\frac{\Gamma^{\mu\alpha}}{l_L^2}\frac{\Gamma^{\nu\beta}}{l_R^2}
$$

    We have

$$\alpha_s^2\sum_iP_i^{qq\rightarrow qq}(x_1,x_2,x_3,x_4)\vert_{x_3=x_B}$$
$$=\frac{1}{16\pi ^2}\frac{x_3x_4}{(x_1+x_2)^3}
[M_tM_t^* \vert_{x_3=x_B}+M_uM_u^* \vert_{x_3=x_B}
 +2M_tM_u^* \vert_{x_3=x_B}]$$

and

$$\alpha_s^2\sum_iP_i^{qq\rightarrow qq}(x_1,x_2,x_3,x_4)\vert_{x_4=x_B}$$
$$=\frac{1}{16\pi ^2}\frac{x_3x_4}{(x_1+x_2)^3}
[M_tM_t^* \vert_{x_4=x_B}+M_uM_u^* \vert_{x_4=x_B} +2M_tM_u^*
\vert_{x_4=x_B}]$$

    Therefore£¬ the recombination is

$$P^{qq\rightarrow q }(x_1,x_B)$$
$$=\int
\sum_iP_i^{qq\rightarrow qq}(x_1,x_2,x_3,x_4)
\delta(x_1+x_2-x_3-x_4)
[\delta(x_3-x_B)+\delta(x_4-x_B)]dx_3dx_4$$
$$=\frac{2}{9}\frac{(2x_1-x_B)^2}{x_1^3}.\eqno(A-3)$$

4.  $q\overline q \rightarrow q\overline q$ (Fig.10).

    Note that $M_t(q\bar q\rightarrow q\bar q)M_t^*(q\bar q\rightarrow q\bar q)=
M_t(qq\rightarrow qq)M_t^*(qq\rightarrow qq)$.

$$ M_sM^*_s\vert_{x_3=x_B}=\frac{g^4}{4}<\frac{2}{9}> Tr[\gamma\cdot p_1  \gamma^\mu
\gamma\cdot p_2 \gamma^\nu] \frac{n^\mu n^\alpha}{l_L^2}\frac{n^\nu
n^\beta}{l_R^2} Tr[\gamma\cdot k  \gamma^\beta
\gamma\cdot l' \gamma^\alpha]  $$

$$ M_tM^*_s\vert_{x_3=x_B}=\frac{g^4}{4}<-\frac{2}{27}> Tr[\gamma\cdot
p_1 \gamma^\mu\gamma\cdot k  \gamma^\beta
\gamma\cdot l' \gamma^\alpha \gamma\cdot p_2  \gamma^\nu ] \frac{\Gamma^{\mu\alpha}}{l_L^2}\frac{n^\nu
n^\beta}{l_R^2} $$

     We have

$$\alpha_s^2\sum_iP_i^{q\overline q \rightarrow q\overline
q}(x_1,x_2,x_3,x_4)\vert_{x_3=x_B} $$
$$=\frac{1}{16\pi
^2}\frac{x_3x_4}{(x_1+x_2)^3} [M_tM_t^* \vert_{x_3=x_B}+ M_sM_s^*
\vert_{x_3=x_B}+ 2M_tM_s^* \vert_{x_3=x_B}]$$

and

$$\alpha_s^2\sum_iP_i^{q\overline q \rightarrow q\overline
q}(x_1,x_2,x_3,x_4)\vert_{x_4=x_B}$$
$$=\frac{1}{16\pi ^2}\frac{x_3x_4}{(x_1+x_2)^3}
[M_tM_t^* \vert_{x_4=x_B} + M_sM_s^* \vert_{x_4=x_B} +2M_tM_s^*
\vert_{x_4=x_B}]$$

    Therefore, the recombination is

$$P^{q \overline q\rightarrow q}(x_1,x_B)=P^{q \overline q\rightarrow \overline q}(x_1,x_B)$$
$$=\int
\sum_iP_i^{q\overline q \rightarrow q\overline
q}(x_1,x_2,x_3,x_4)\delta(x_1+x_2-x_3-x_4)\delta(x_3-x_B)dx_3dx_4$$
$$=\frac{1}{108}\frac{(2x_1-x_B)^2(6x_1^2+x_1x_B+3x_B^2)}{x_1^5}.\eqno(A-4)$$

5. ${q \overline q \rightarrow GG}$ (Fig.11).

$$ M_tM^*_t\vert_{x_3=x_B}
=\frac{g^4}{4}<\frac{16}{27}> Tr[\gamma\cdot p_1  \gamma^\mu
\frac{\gamma\cdot l_L}{(l_L)^2} \gamma^\alpha \gamma\cdot p_2
\gamma^\beta  \frac{\gamma\cdot l_R}{l_R^2} \gamma^\nu]$$
$$[\delta^{ij}-\frac{k^ik^j}{\left\vert\vec{k}\right\vert^2}]
[\delta^{lm}-\frac{l'^ll'^m}{\left\vert\vec{l'}\right\vert^2}],
 $$

$$ M_uM^*_u\vert_{x_3=x_B}
=\frac{g^4}{4}<\frac{16}{27}> Tr[\gamma\cdot p_1  \gamma^\alpha
\frac{\gamma\cdot l_L}{(l_L)^2} \gamma^\mu \gamma\cdot p_2
\gamma^\nu  \frac{\gamma\cdot l_R}{l_R^2} \gamma^\beta]$$
$$[\delta^{ij}-\frac{k^ik^j}{\left\vert\vec{k}\right\vert^2}]
[\delta^{lm}-\frac{l'^ll'^m}{\left\vert\vec{l'}\right\vert^2}],
 $$

$$ M_sM^*_s\vert_{x_3=x_B}=\frac{g^4}{4}<\frac{4}{3}> Tr[\gamma\cdot p_1  \gamma^\kappa
\gamma\cdot p_2  \gamma^\lambda] \frac{n^\kappa n^\rho}{(l_L\cdot
n)^2}\frac{n^\lambda n^\sigma}{(l_R\cdot n)^2}
C^{\rho\mu\alpha}C^{\sigma\nu\beta} $$
$$[\delta^{ij}-\frac{k^ik^j}{\left\vert\vec{k}\right\vert^2}]
[\delta^{lm}-\frac{l'^ll'^m}{\left\vert\vec{l'}\right\vert^2}],
$$

$$ M_tM^*_u\vert_{x_3=x_B}
=\frac{g^4}{4}<-\frac{2}{27}> Tr[\gamma\cdot p_1  \gamma^\mu
\frac{\gamma\cdot l_L}{(l_L)^2} \gamma^\alpha \gamma\cdot p_2
\gamma^\nu  \frac{\gamma\cdot l_R}{l_R^2} \gamma^\beta]$$
$$[\delta^{ij}-\frac{k^ik^j}{\left\vert\vec{k}\right\vert^2}]
[\delta^{lm}-\frac{l'^ll'^m}{\left\vert\vec{l'}\right\vert^2}],
$$

$$ M_tM^*_s\vert_{x_3=x_B}
=\frac{-ig^4}{4}<\frac{2i}{3}> Tr[\gamma\cdot p_1  \gamma^\mu
\frac{\gamma\cdot l_L}{(l_L)^2} \gamma^\alpha \gamma\cdot p_2
\gamma^\lambda ] \frac{n^{\lambda} n^{\sigma}}{(l_R\cdot
n)^2}C^{\sigma\nu\beta}$$
$$[\delta^{ij}-\frac{k^ik^j}{\left\vert\vec{k}\right\vert^2}]
[\delta^{lm}-\frac{l'^ll'^m}{\left\vert\vec{l'}\right\vert^2}],
 $$

$$ M_uM^*_s\vert_{x_3=x_B}=\frac{-ig^4}{4}<\frac{-2i}{3}> Tr[\gamma\cdot p_1  \gamma^\alpha
\frac{\gamma\cdot l_L}{(l_L)^2} \gamma^\mu \gamma\cdot p_2
\gamma^\lambda ] \frac{n^{\lambda} n^{\sigma}}{(l_R\cdot
n)^2}C^{\sigma\nu\beta}$$
$$[\delta^{ij}-\frac{k^ik^j}{\left\vert\vec{k}\right\vert^2}]
[\delta^{lm}-\frac{l'^ll'^m}{\left\vert\vec{l'}\right\vert^2}],
$$

    We have

$$\alpha_s^2\sum_iP_i^{q\overline q \rightarrow
GG}(x_1,x_2,x_3,x_4)\vert_{x_3=x_B} =\frac{1}{16\pi
^2}\frac{x_3x_4}{(x_1+x_2)^3} [M_tM_t^* \vert_{x_3=x_B}+M_uM_u^*
\vert_{x_3=x_B}$$
$$+ M_sM_s^* \vert_{x_3=x_B}
 +2M_tM_u^* \vert_{x_3=x_B}+2M_tM_s^* \vert_{x_3=x_B}+ 2M_uM_s^*
\vert_{x_3=x_B}]$$

and

$$\alpha_s^2\sum_iP_i^{q\overline q \rightarrow
GG}(x_1,x_2,x_3,x_4)\vert_{x_4=x_B} =\frac{1}{16\pi
^2}\frac{x_3x_4}{(x_1+x_2)^3} [M_tM_t^* \vert_{x_4=x_B}+M_uM_u^*
\vert_{x_4=x_B}$$
$$+ M_sM_s^* \vert_{x_4=x_B}
 +2M_tM_u^* \vert_{x_4=x_B}+2M_tM_s^* \vert_{x_4=x_B}+ 2M_uM_s^*
\vert_{x_4=x_B}]$$

   Therefore, the recombination is

$$P^{q \overline q\rightarrow G}(x_1,x_B)$$
$$=\int
\sum_iP_i^{q\overline q \rightarrow GG
}(x_1,x_2,x_3,x_4)\delta(x_1+x_2-x_3-x_4)[ \delta(x_3-x_B)+\delta(x_4-x_B)]dx_3dx_4$$
$$=\frac{4}{27}\frac{(2x_1-x_B)(18x_1^2-9x_1x_B+4x_B^2)}{x_1^3x_B}
.\eqno(A-5)$$

6.  ${qG \rightarrow qG}$ (Fig.12).

$$ M_tM^*_t\vert_{x_3=x_B}
=\frac{g^4}{4}<\frac{1}{2}> Tr[\gamma\cdot p_1  \gamma^\kappa
\gamma\cdot k
\gamma^\lambda]\frac{\Gamma^{\kappa\rho}}{(l_L)^2}\frac{\Gamma^{\lambda\sigma}}{(l_R)^2}C^{\rho\phi\alpha}C^{\sigma\beta\chi}\delta_\perp^{rs}
[\delta^{lm}-\frac{l'^ll'^m}{\left\vert\vec{l'}\right\vert^2}]
$$

$$ M_uM^*_u\vert_{x_3=x_B}
=\frac{g^4}{4}<\frac{2}{9}> Tr[\gamma\cdot p_1 \gamma^\alpha
\frac{\gamma\cdot l_L}{(l_L)^2} \gamma^\phi\gamma\cdot k \gamma^\chi
\frac{\gamma\cdot l_R}{(l_R)^2} \gamma^\beta]\delta_\perp^{rs}
[\delta^{lm}-\frac{l'^ll'^m}{\left\vert\vec{l'}\right\vert^2}]
$$

$$ M_sM^*_s\vert_{x_3=x_B}=\frac{g^4}{4}<\frac{2}{9}> Tr[\gamma\cdot p_1 \gamma^\phi
\frac{\gamma\cdot n}{2l_L\cdot n} \gamma^\alpha\gamma\cdot
k\gamma^\beta \frac{\gamma\cdot n}{2l_R\cdot n}
\gamma^\chi]\delta_\perp^{rs}
[\delta^{lm}-\frac{l'^ll'^m}{\left\vert\vec{l'}\right\vert^2}]
$$

$$ M_tM^*_u\vert_{x_3=x_B}=\frac{ig^4}{4}<\frac{i}{4}> Tr[
\gamma\cdot p_1 \gamma^\kappa\gamma \cdot k\gamma^\chi
 \frac{\gamma\cdot l_R}{l_R^2}
\gamma^\beta]\frac{\Gamma^{\kappa\rho}}{l_L^2}
C^{\rho\phi\alpha}\delta_\perp^{rs}
[\delta^{lm}-\frac{l'^ll'^m}{\left\vert\vec{l'}\right\vert^2}]
$$

$$ M_tM^*_s\vert_{x_3=x_B}
=\frac{ig^4}{4}<-\frac{i}{4}> Tr[
\gamma\cdot p_1 \gamma^\kappa \gamma\cdot k\gamma\cdot \beta\frac{\gamma\cdot n}{2l_R\cdot n}
\gamma^\chi]
\frac{\Gamma^{\kappa\rho}}{(l_L)^2}C^{\rho\phi\alpha}\delta_\perp^{rs}
[\delta^{lm}-\frac{l'^ll'^m}{\left\vert\vec{l'}\right\vert^2}]
$$

$$ M_uM^*_s\vert_{x_3=x_B}
=\frac{g^4}{4}<-\frac{1}{36}> Tr[\gamma\cdot p_1 \gamma^\alpha
\frac{\gamma\cdot l_L}{(l_L)^2}\gamma^\phi \gamma\cdot k
\gamma^\beta \frac{\gamma\cdot n}{2l_R\cdot n}
\gamma^\chi
]\delta_\perp^{rs}
[\delta^{lm}-\frac{l'^ll'^m}{\left\vert\vec{l'}\right\vert^2}]
$$

    We have

$$\alpha_s^2\sum_iP_i^{qG \rightarrow
qG}(x_1,x_2,x_3,x_4)\vert_{x_3=x_B} =\frac{1}{16\pi
^2}\frac{x_3x_4}{(x_1+x_2)^3} [M_tM_t^* \vert_{x_3=x_B}+M_uM_u^*
\vert_{x_3=x_B}$$
$$+ M_sM_s^* \vert_{x_3=x_B}
 +2M_tM_u^* \vert_{x_3=x_B}+2M_tM_s^* \vert_{x_3=x_B}+ 2M_uM_s^*
\vert_{x_3=x_B}]$$

and

$$\alpha_s^2\sum_iP_i^{qG \rightarrow
qG}(x_1,x_2,x_3,x_4)\vert_{x_4=x_B} =\frac{1}{16\pi
^2}\frac{x_3x_4}{(x_1+x_2)^3} [M_tM_t^* \vert_{x_4=x_B}+M_uM_u^*
\vert_{x_4=x_B}$$
$$+ M_sM_s^* \vert_{x_4=x_B}
 +2M_tM_u^* \vert_{x_4=x_B}+2M_tM_s^* \vert_{x_4=x_B}+ 2M_uM_s^*
\vert_{x_4=x_B}]$$

     Therefore, the recombination is

$$P^{qG\rightarrow G}(x_1,x_B)$$
$$=\int
\sum_iP_i^{qG \rightarrow
qG}(x_1,x_2,x_3,x_4)\delta(x_1+x_2-x_3-x_4) \delta(x_4-x_B)dx_3dx_4$$
$$=\frac {1}{288}\frac {
(2x_1-x_B)^2
(304x_1^2-202x_1x_B+79x_B^2)}{x_1^4x_B},
,\eqno(A-6)$$

and

$$P^{qG\rightarrow q}(x_1,x_B)$$
$$=\int
\sum_iP_i^{qG \rightarrow G
q}(x_1,x_2,x_3,x_4)\delta(x_1+x_2-x_3-x_4) \delta(x_3-x_B)dx_3dx_4$$
$$=\frac {1}{288}\frac {
(2x_1-x_B)
(140x_1^2-52x_1x_B+65x_B^2)}{x_1^4}. \eqno(A-7)$$

\newpage
\noindent {\bf Figure Captions}

\noindent Fig. 1 The diagrams contributing to a new modified Altarelli-Parisi
equation with gluon recombination. Here the shaded part implies the
correlation of gluons at short-distance and $"\times"$ means the probing place.

\noindent Fig. 2 The two-gluon-to-two-gluon subprocess in recombination
function.

\noindent Fig. 3 A $t$-channel diagram for $GG\rightarrow G$.

\noindent Fig. 4 A $s$-channel diagram for $GG\rightarrow G$.

\noindent Fig. 5 A $t$-channel diagram for $GG\rightarrow q$.

\noindent Fig. 6 A dominant splitting process in the DLLA, where
$x_B\ll x_1$ leads to a strong order in $x$.

\noindent Fig. 7 All channels for $GG\rightarrow GG$.

\noindent Fig. 8 All channels for $GG\rightarrow q \overline q$.

\noindent Fig. 9 All channels for  $qq\rightarrow qq$.

\noindent Fig. 10 All channels for $q\overline q \rightarrow q\overline q$.

\noindent Fig. 11 All channels for ${q \overline q \rightarrow GG}$.

\noindent Fig. 12 All channels for ${qG \rightarrow qG}$.

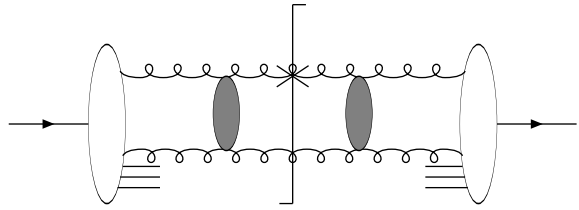
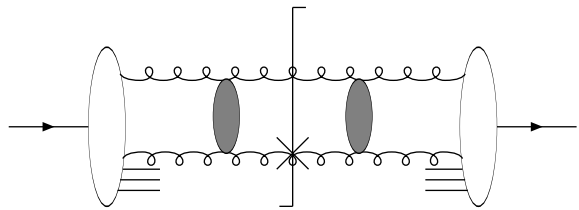
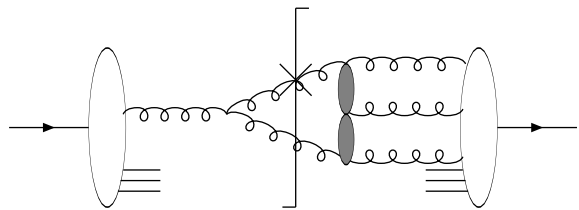
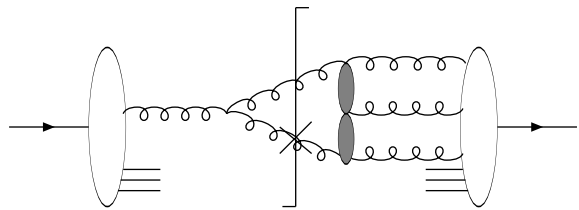
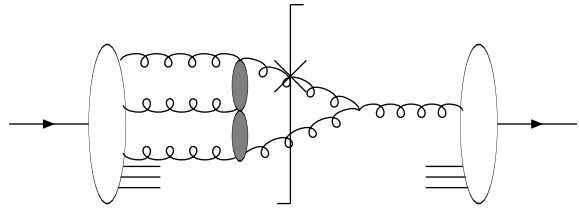
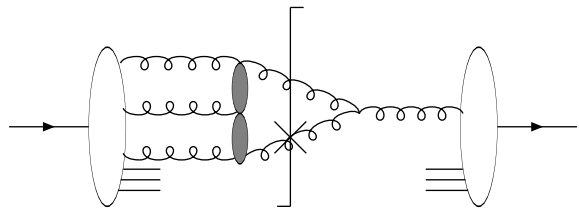
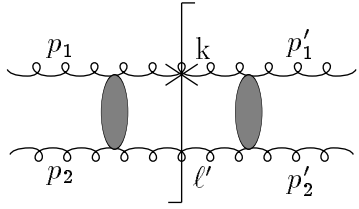
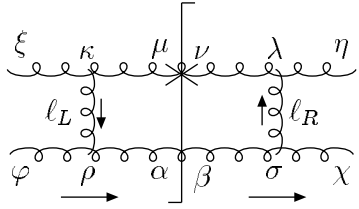
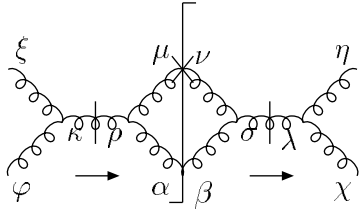
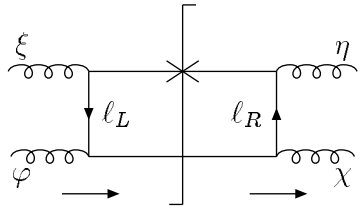
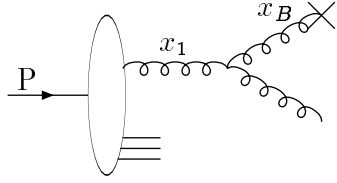
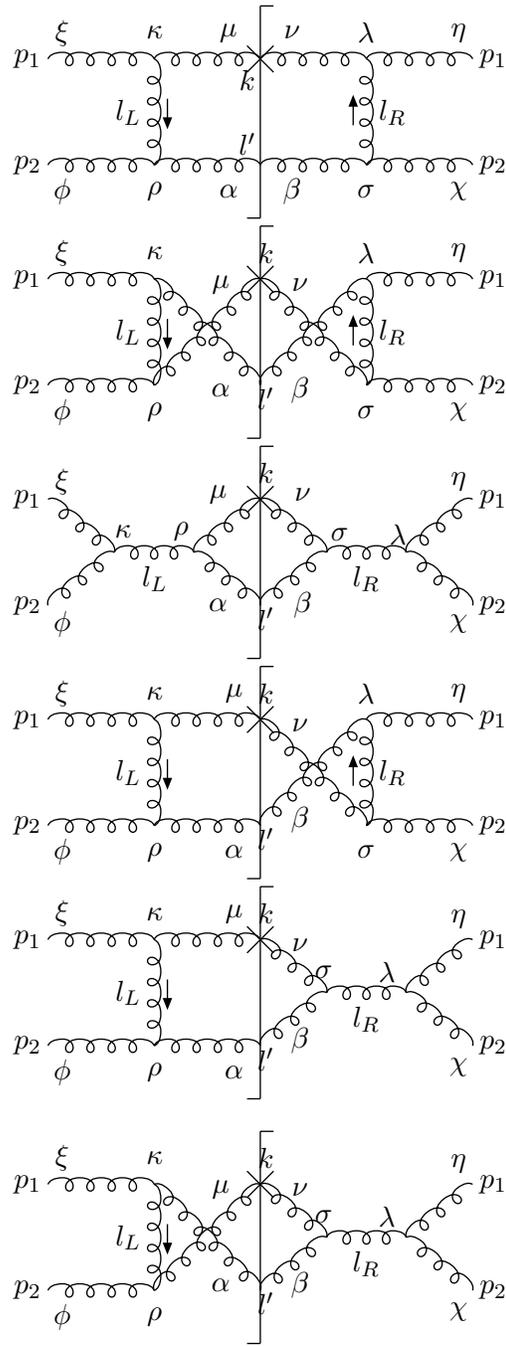
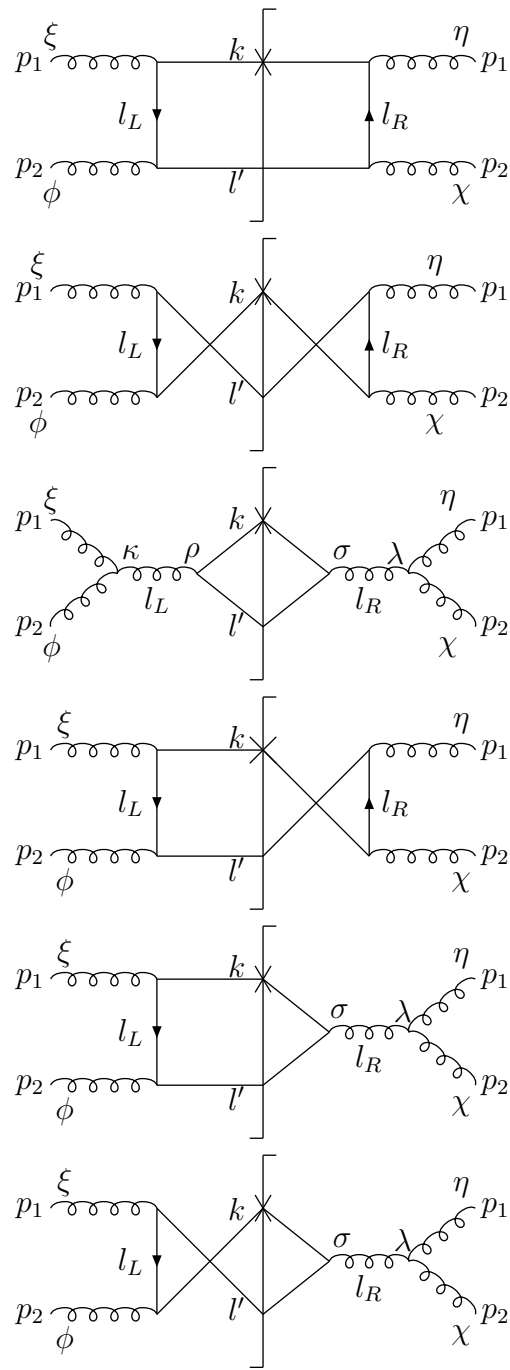
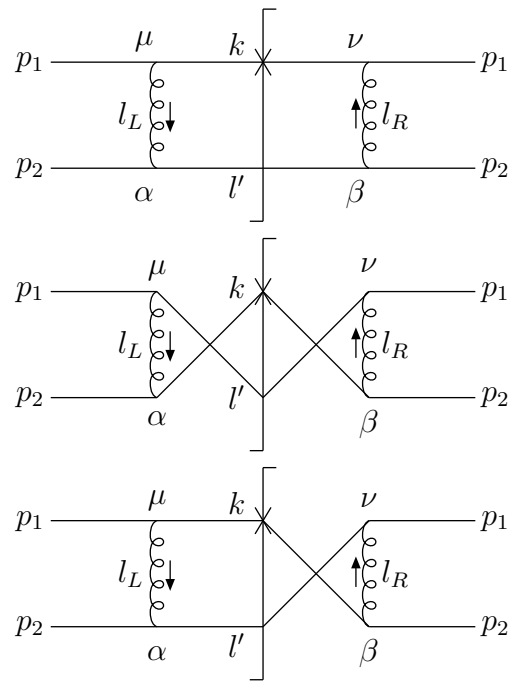
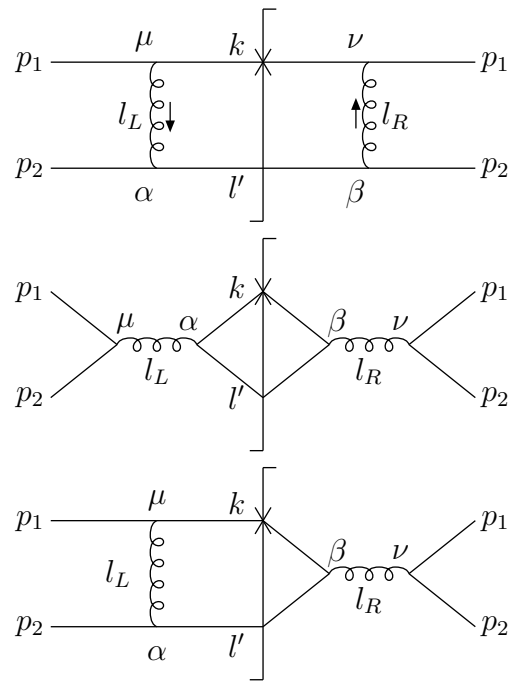
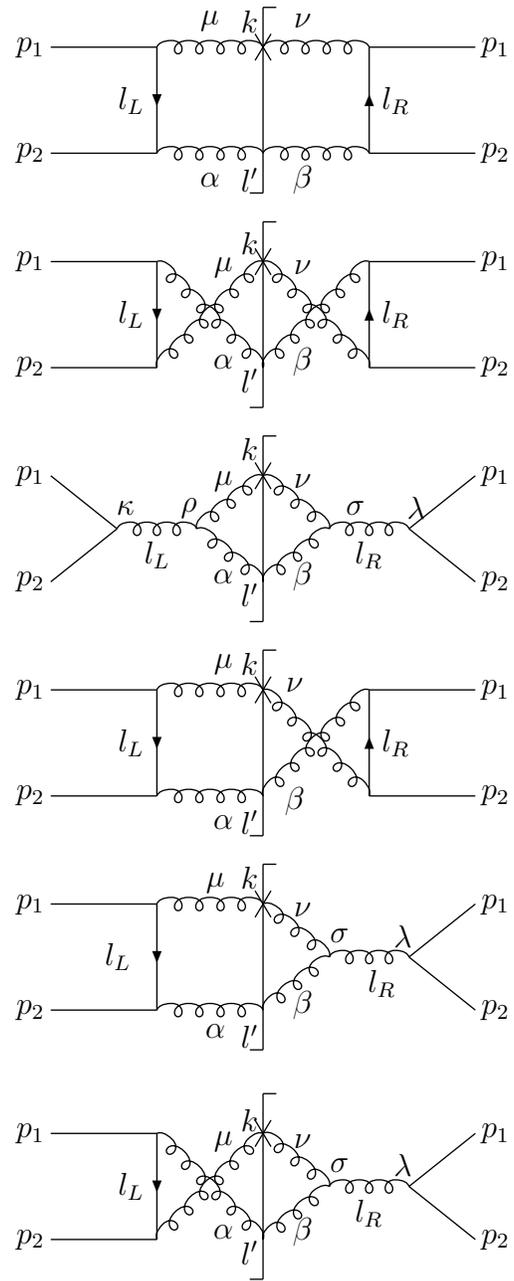
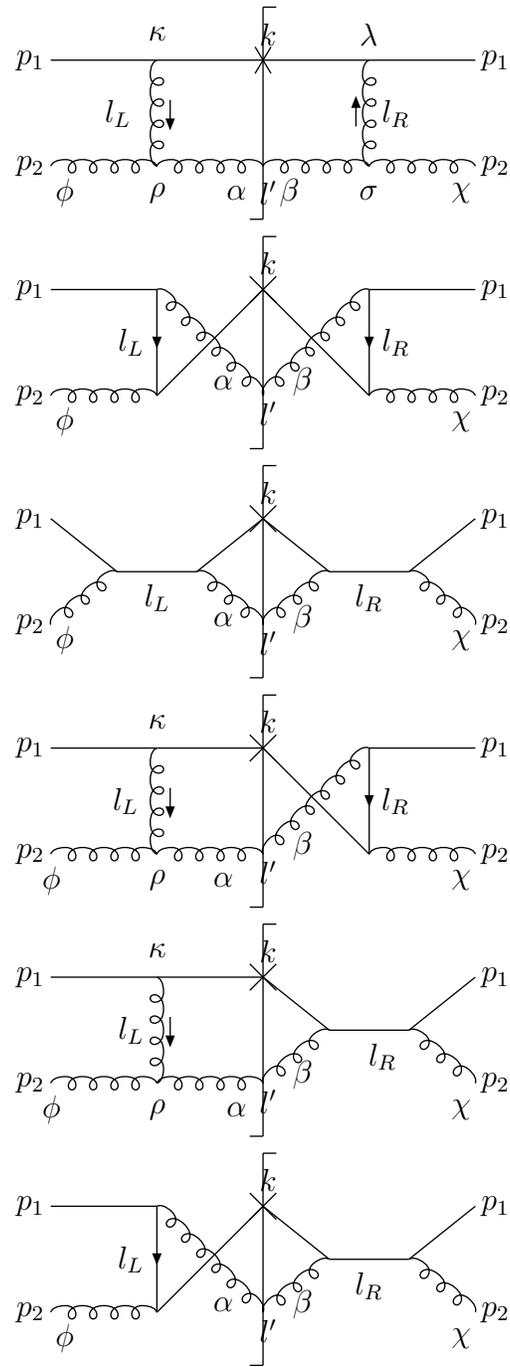
\end{document}